\documentclass[twocolumn,aps,prl,showpacs,superscriptaddress,tightenlines]{revtex4-1}
\usepackage{amsmath}
\usepackage{graphicx}
\usepackage{color}
\usepackage{amsfonts}
\usepackage{txfonts}
\usepackage[colorlinks,citecolor=blue]{hyperref}
\usepackage{multirow}

\def\narrowtext{\par\global\columnwidth20.5pc
\global\hsize\columnwidth\global\linewidth\columnwidth
\global\displaywidth\columnwidth}

\begin{document}
\title{Collective Quantum Entanglement in Molecular Cavity Optomechanics}

\author{Jian Huang}
\affiliation{Department of Physics, City University of Hong Kong, Kowloon, Hong Kong SAR}

\author{Dangyuan Lei}
\affiliation{Department of Materials Science and Engineering, City University of Hong Kong, Kowloon, Hong Kong SAR}

\author{Girish S. Agarwal}
\affiliation{Institute for Quantum Science and Engineering, Department of Biological and Agricultural Engineering, Texas A$\&$M University, College Station, Texas 77843, USA}

\author{Zhedong Zhang}
\email{zzhan26@cityu.edu.hk}
\affiliation{Department of Physics, City University of Hong Kong, Kowloon, Hong Kong SAR}
\affiliation{City University of Hong Kong, Shenzhen Research Institute, Shenzhen, Guangdong 518057, China}

\begin{abstract}
We propose an optomechanical scheme for reaching quantum entanglement in vibration polaritons. The system involves $N$ molecules, whose vibrations can be fairly entangled with plasmonic cavities. We find that the vibration-photon entanglement can exist at room temperature and is robust against thermal noise. We further demonstrate the quantum entanglement between the vibrational modes through the plasmonic cavities, which shows a delocalized nature and an incredible enhancement with the number of molecules. The underlying mechanism for the entanglement is attributed to the strong vibration-cavity coupling which possesses collectivity. Our results provide a molecular optomechanical scheme which offers a promising platform for the study of noise-free quantum resources and macroscopic quantum phenomena.
\end{abstract}
\maketitle
\narrowtext

\emph{Introduction.}---Recent efforts have been largely devoted to studying the properties of strong coupling of molecules to light, which forms a new excitation known as the molecular polariton (MP)~\cite{Guebrou2012,Spano2015,Flick2017,Su2020}. The molecular polaritons are of a wide spectrum, ranging from far infrared to ultraviolet regimes~\cite{ Xiang2018,Ribeiro2018,Zhang2019a,Zhang2023}. It has been shown an incredible modification of exciton dynamics and reaction kinetics via the strong coupling to microcavities~\cite{Hutchison2012,Coles2014,Thomas2016,Dunkelberger2016,Herrera2016,Kowalewski2016,Galego2016,Martinez-Martinez2018,Li2021}. Moreover, the polaritons demonstrated their quantum nature prominently in the optomechanical scheme, including Bose condensations, squeezed states, and entanglement~\cite{Aspelmeyer2014,Metcalfe2014}. The strong molecule-photon coupling therefore led to a prosperous feature of several fields such as molecular electronics~\cite{Joachim2000,Cuevas2010,Xiang2016}, lasers~\cite{Kena-Cohen2010}, and optomechanics~\cite{Aspelmeyer2014,Shalabney2015,Pino2015,Xiang2024}.

 Molecular cavity optomechanics~\cite{Roelli2016,Schmidt2016,Benz2016,Dezfouli2017,Lombardi2018,Neuman2019,Zhang2020,Esteban2022,Jakob2023,Koner2023}, known as a combination of MPs and optomechanics, is a powerful subject when exploring the spectroscopy and metrology, in which the quantum entanglement may play an important role. So far, the quantum entanglement has emerged in a variety of systems, e.g., quantum liquid~\cite{Zhou2017,Savary2017}, ferromagnetic materials~\cite{Ghosh2003,Li2018,Zhang2019b}, and trapped ions~\cite{Leibfried2003}. Thanks to the advancements of cooling technique and material synthesis~\cite{Wilson-Rae2007,Marquardt2007,Chan2011,Teufel2011a,Barzanjeh2022}, cavity optomechanical model has become a promising platform to achieve prominent entanglement~\cite{Mancini2002,Pirandola2006,Vitali2007prl,Vitali2007jpa,Paternostro2007,Genes2008,Hartmann2008,Borkje2011,Abdi2012,Tian2013,Wang2013,Palomaki2013,Flayac2014,Liao2014,Ho2018,Barzanjeh2019,Jiao2020,Lai2022,Huang2022,Riedinger2018,Ockeloen-Korppi2018,Kotler2021,Lepinay2021}. The cavity optomechanics may enable a capability of assembling entanglement robust to external noise and fluctuations. This is fairly important for an extensive study of complex materials including magnons and molecules that contain much richer degrees of freedom than atoms. Recent progresses reported the entanglement in ferromagnetic bulks against the thermal noise and disorder~\cite{Li2018,Zhang2019b}. For molecular systems, elaborate experiments demonstrated the collective coupling to cavity photons, yielding the Rabi oscillation that led to an incredibly enhanced up-conversion efficiency and a modification of matter phases~\cite{Chen2021,Xomalis2021}. In this regard, the long-range coherence over many molecules arising from the collectivity turns out to be of a central aspect, so as to combat the local fluctuations. This however is still an open issue.

 The collective molecule-cavity interaction essentially brings out a $\sqrt{N}$ enhancement of the optomechanical coupling~\cite{Agarwal1984,Zou2024}. This used to be a hard task in the cavity optomechanics. In this Letter, we study the quantum entanglement in the molecular cavity optomechanical systems. The $\sqrt{N}$ scaling is shown to be a key for reaching the entanglement between molecular vibrations. We find that, under feasible parameter regimes, the cavity-vibration entanglement can survive at room temperature and is highly robust to the thermal noise. Moreover, our results reveal a delocalized nature of the vibration-vibration entanglement, evident by the unusual scaling of the logarithmic negativity against the size of molecular sample. This further leads to a remarkable enhancement when having a large number of molecules in the sample. Our work  offers a new paradigm for generating quantum resources and for studying the quantum information processing using molecular optomechanical systems.

%%%%%%%%%%%%%%%%%%%%%%%%%%%%%%%
\begin{figure}[tbp]
\center
\includegraphics[width=0.47 \textwidth]{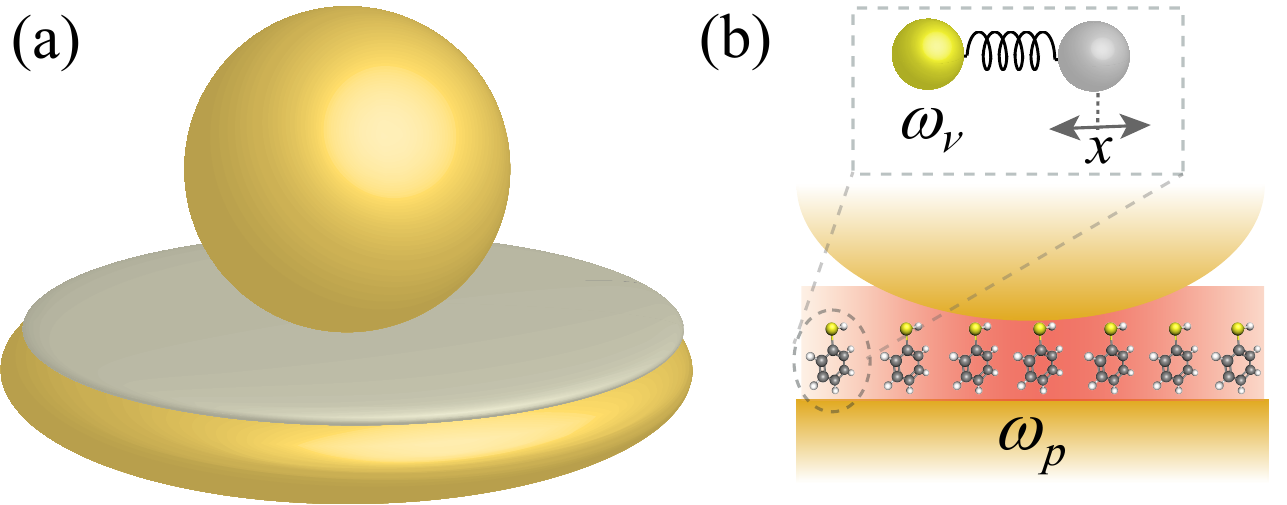}
\caption{(a) Schematic of the nanoparticle-on-mirror configuration. From top to bottom, they are metallic nanoparticle, self-assembled monolayer of thiophenol molecules, and metallic substrate, respectively.  (b)  Schematic of the molecular cavity optomechanical system. Plasmonic cavity mode (with resonance frequency $\omega_{p}$) is coupled to $N$ molecular vibrational modes (resonance frequency $\omega_{v}$) via the optomechanical interactions. Here the vibrational  mode is depicted as two masses linked by a spring.}
\label{Fig1}
\end{figure}
%%%%%%%%%%%%%%%%%%%%%%%%%%%%%%

\emph{Molecular model and Langevin equations.}---We consider a  molecular cavity optomechanical system consisting of a plasmonic cavity and $N$ identical molecules.  The plasmonic system can be realized in the nanoparticle-on-mirror configuration~\cite{Mubeen2012} or the scanning-tunnelling microscope tip on metallic substrate~\cite{Zhang2013}, where $N$ organic molecules are sandwiched in the gap between the nanostructure and the metallic substrate (see Fig.~\ref{Fig1}). Molecular optomechanics possesses two notable advantages: (i) high molecule-photon coupling from the tightly-confined mode volume of cavities; (ii) high vibration frequency where the thermalization is suppressed at room temperature, that is significant for quantum applications. In this vein, the molecular optomechanical system provide a promising platform for investigating quantum entanglement.
In the low-excitation-number case,  the vibrational mode of  each molecule can be approximately described by a harmonic oscillator with resonance frequency $\omega_{v}$ and a normal coordinate $x=x_{\text{zpf}}(b^{\dagger}+b)$, where $x_{\text{zpf}}$ is the zero-point amplitude and $b$ $(b^{\dagger})$ is the annihilation (creation) operator. We assume that the cavity resonance frequency $\omega_{p}$ is off-resonant with the optically allowed electronic transitions of the molecule. The vibration-cavity coupling is thus  parametric. The molecular vibration thus modulates the cavity resonance frequency, i.e., $\omega_{p}(x)\approx\omega_{p}-\tilde{g}_{v}x$, where $\tilde{g}_{v}=\omega_{p}(\partial \alpha/\partial x)/(\varepsilon_{0}V_{p})$, with $\alpha$, $\varepsilon_{0}$, and $V_{p}$ being the respective polarizability of the molecule, the vacuum permittivity, and the volume of the cavity~\cite{Roelli2016}. The Hamiltonian of the system is of the form ($\hbar=1$)
\begin{equation}
 \label{Hamit}
 H=\omega_{p}a^{\dagger}a+\sum_{j=1}^{N}[\omega_{v}b_{j}^{\dagger}b_{j}+g_{v}a^{\dagger}a(b_{j}^{\dagger}+b_{j})]+(\Omega a^{\dagger} e^{-i\omega _{l}t} +\mathrm{H.c.}),  \\
\end{equation}
where $a$ $(a^{\dagger})$ and $b_{j}$ $(b^{\dagger}_{j})$ are the annihilation (creation) operators of the cavity mode and the vibrational mode of the $j$th molecule, respectively. $g_{v} = -\tilde{g}_{v}x_{\text{zpf}}$ is the molecular optomechanical coupling constant~\cite{SMaterial};  $\Omega$ and $\omega_{l}$ are the amplitude and frequency of the pump field, respectively. With the effective mass of the vibrations, one finds $g_{v}= -\omega_{p} R_{v}\sqrt{\hbar/2 m \omega_{v}}/(\varepsilon_{0}V_{p})$,  where $R_{v}=\partial \alpha/\partial x$ and $\alpha$ is the Raman polarizability. As an estimation~\cite{Roelli2016}, $\omega_{p}/2\pi=333$ THz, $\omega_{v}/2\pi=29.9$ THz, $R_{v}^2/m \approx 3\times10^{-10}\varepsilon^{2}_{0}$ {\AA}$^{4}$amu$^{-1}$, and  $V_{p}=1.5\times10^{-6}$ $\mu$m$^{3}$, which yield $-g_{v}/2\pi\approx21$ GHz. In recent studies, for both theory~\cite{Roelli2016,Schmidt2016,Esteban2022} and experiments~\cite{Benz2016,Zhang2013,Lombardi2018}, the parameters were taken as $-g_{v}/2\pi\sim10$ GHz-$100$ GHz and $\omega_{v}/2\pi\sim6$ THz-$48$ THz ($\approx200$ cm$^{-1}$-$1600$ cm$^{-1}$).

We consider a large number of molecules in the cavity, i.e., $N\gg1$, that is always feasible in labs. Our motivation is to study the entanglement between the molecules. To this end, one defines the two collective vibrational modes $B_{1}=\sum_{j=1}^{M}b_{j}/\sqrt{M}$ and $B_{2}=\sum_{j=M+1}^{N}b_{j}/\sqrt{N-M}$ (with $[B_l, B^{\dagger}_{l^{\prime}}]=\delta_{ll^{\prime}}$)~\cite{Sun2003,Emary2003}. The Hamiltonian $H$ in the rotating frame of the drive frequency $\omega_{l}$  can be rewritten as
\begin{equation}
H_{I}=\Delta_{p}a^{\dagger}a+\sum_{l=1}^{2}[\omega_{v}B_{l}^{\dagger}B_{l}+g_{l}a^{\dagger}a(B_{l}^{\dagger}+B_{l})]+(\Omega a^{\dagger}+\mathrm{H.c.}),  \\
\end{equation}
where $\Delta_{p}=\omega_{p}-\omega_{l}$  is the driving detuning, and $g_{1}=g_{v}\sqrt{M}$ ($g_{2}=g_{v}\sqrt{N-M}$) is the collective optomechanical coupling strength between the cavity mode $a$ and the collective vibrational mode $B_{1}$ ($B_{2}$).

With a strong drive, the dynamics of this system can be linearized  by writing each operator into $o=\langle o\rangle_{\text{ss}}+\delta o$, where $\langle o\rangle_{\text{ss}}$ and $\delta o$ denote the steady-state mean and the quantum fluctuation, respectively. The linearized quantum Langevin equations are thus obtained
\begin{eqnarray}
\label{linzedlang}
\delta \dot{a} &=&-( i\Delta +\kappa) \delta a-i\sum_{l=1}^{2}G_{l}(\delta B_{l}^{\dagger }+\delta B_{l}) +\sqrt{2\kappa }a_{\text{in}},  \nonumber \\
\delta \dot{B}_{1} &=&-( i\omega_{v} +\gamma_{1}) \delta B_{1}-i(G_{1}^{\ast}\delta a+G_{1}\delta a^{\dagger})+\sqrt{2\gamma}B_{1,\text{in}}, \nonumber \\
\delta \dot{B}_{2} &=&-(i\omega_{v} +\gamma_{2}) \delta B_{2}-i(G_{2}^{\ast }\delta a+G_{2}\delta a^{\dagger})+\sqrt{2\gamma }B_{2,\text{in}},\nonumber \\
\end{eqnarray}
where $\kappa$ and $\gamma_{l=1,2}$ are the decay rates of the cavity mode and collective vibrational modes,  respectively; $\Delta=\Delta_{p}+\sum_{l=1}^{2} g_{l}(\langle B_{l}\rangle_{\text{ss}}+\langle B_{l}\rangle^{\ast}_{\text{ss}})$ is the normalized driving detuning; $G_{1}=\sqrt{M}G_{v}$ and  $G_{2}=\sqrt{N-M}G_{v}$ ($G_{v}=g_{v}\langle a\rangle_{\text{ss}}$) are the linearized collective (single-photon) optomechanical coupling strengths. The collectivity of $G_l$ in the molecular cavity optomechanics may lead to the ultrastrong-coupling regime, while it used to be weak in conventional optomechanics. Thus the term $G_l \delta B_l^{\dagger}$ in Eq.~(\ref{linzedlang}) resulting from the counter-rotating-wave effect cannot be neglected, and it is the key component for generating entanglement. Elaborate analysis will be presented later on. The steady-state means are
\begin{equation}
\label{steady}
\langle a\rangle_{\text{ss}}=\frac{-i\Omega}{i\Delta +\kappa}, \hspace{0.7 cm} \langle B_{l}\rangle_{\text{ss}}=\frac{-ig_{l}\vert\langle a\rangle_{\text{ss}}\vert^{2} }{i\omega_{v}+\gamma_{l}}.
\end{equation}
The noise operators $a_{\textrm{in}}$, $B_{1,\textrm{in}}=\sum_{j=1}^{M}b_{j\text{,in}}/\sqrt{M}$,  and  $B_{2,\textrm{in}}=\sum_{j=M+1}^{N}b_{j\text{,in}}/\sqrt{N-M}$ are characterized by the nonzero correlation functions~\cite{Gardiner2013},
$\langle a_{\textrm{in}}(t) a_{\textrm{in}}^{\dagger}(t^{\prime})\rangle=\delta(t-t^{\prime})$,
$\langle B_{l,\textrm{in}}(t) B_{l,\textrm{in}}^{\dagger}(t^{\prime})\rangle=(\bar{n}_{l}+1)\delta(t-t^{\prime})$, and
$\langle B_{l,\textrm{in}}^{\dagger}(t)B_{l,\textrm{in}}(t^{\prime})\rangle=\bar{n}_{l}\delta(t-t^{\prime})$, with $\bar{n}_{l}$ being the thermal phonon number for the vibrational mode $B_{l}$.

 By defining the vectors of the quadrature fluctuation operators $\mathbf{u}(t)=(\delta{X}_{B_{1}},\delta{Y}_{B_{1}},\delta{X}_{B_{2}},\delta{Y}_{B_{2}},\delta{X}_{a},\delta{Y}_{a})^{T}$ and the noise operators $\mathbf{N}(t)=\sqrt{2}(\sqrt{\gamma_{1}}X_{B_{1}}^{\text{in}},\sqrt{\gamma_{1}}Y_{B_{1}}^{\text{in}},\sqrt{\gamma_{2}}X_{B_{2}}^{\text{in}},\sqrt{\gamma_{2}}Y_{B_{2}}^{\text{in}},\sqrt{\kappa}X_{a}^{\text{in}},\sqrt{\kappa}Y_{a}^{\text{in}})^{T}$, with $\delta X_{o}=(\delta o^{\dagger}+\delta o)/\sqrt{2}$, $\delta Y_{o}=i(\delta o^{\dagger}-\delta o)/\sqrt{2}$, $X_{o}^{\text{in}}=(o_{\text{in}}^{\dagger}+o_{\text{in}})/\sqrt{2}$, and $Y_{o}^{\text{in}}=i(o_{\text{in}}^{\dagger}-o_{\text{in}})/\sqrt{2}$ for $o_{\text{in}}=a_{\text{in}}$ and $B_{l,\text{in}}$, the Eqs.~(\ref{linzedlang}) can be expressed concisely as $\mathbf{\dot{u}}(t)=\mathbf{Au}(t)+\mathbf{N}(t)$, where the drift matrix $\mathbf{A}$ is given by
\begin{equation}
\mathbf{A}=\left(
\begin{array}{cccccc}
-\gamma_{1} & \omega_{v} & 0 & 0 & 0 & 0 \\
-\omega_{v} & -\gamma_{1} & 0 & 0 & -2G_{1}^{\text{Re}} & -2G_{1}^{\text{Im}} \\
0 & 0 & -\gamma_{2} & \omega_{v} & 0 & 0 \\
0 & 0 & -\omega_{v} & -\gamma_{2} &  -2G_{2}^{\text{Re}} & -2G_{2}^{\text{Im}}\\
2G_{1}^{\text{Im}} & 0 & 2G_{2}^{\text{Im}}& 0 & -\kappa & \Delta \\
-2G_{1}^{\text{Re}} & 0 & -2G_{2}^{\text{Re}} & 0 & -\Delta & -\kappa
\end{array}
\right).
\end{equation}
Here $G_{l=1,2}^{\text{Re}}$ ($G_{l}^{\text{Im}}$) is the real (imaginary) part of $G_{l}$. This linearized system can achieve stability only if the real parts of the eigenvalues of $\mathbf{A}$ are negative, as delineated by the Routh-Hurwitz criterion~\cite{Gradstein2014}.

To explore the steady-state quantum entanglement, we introduce the covariance matrix $\mathbf{V}$,  which is characterized by the matrix elements as $\mathbf{V}_{ij}=\frac{1}{2}[\langle \mathbf{u}_{i}(\infty) \mathbf{u}_{j}(\infty) \rangle +\langle \mathbf{u}_{j}( \infty) \mathbf{u}_{i}(\infty )\rangle]$ for $i,j=1\text{-}6$. The covariance matrix $\mathbf{V}$ is determined by the Lyapunov equation  $\mathbf{A}\mathbf{V}+\mathbf{V}\mathbf{A}^{T}=-\mathbf{Q}$~\cite{Vitali2007prl},  where $\mathbf{Q}=\mathrm{diag} \{(2\bar{n}_{1}+1)\gamma_{1},(2\bar{n}_{1}+1)\gamma_{1},(2\bar{n}_{2}+1)\gamma_{2},(2\bar{n}_{2}+1)\gamma_{2},\kappa,\kappa\}$ represents the diffusion matrix.

%%%%%%%%%%%%%%%%%%%%%%%%%%%%%%
\begin{figure}[tbp]
\centering
\includegraphics[width=0.48 \textwidth]{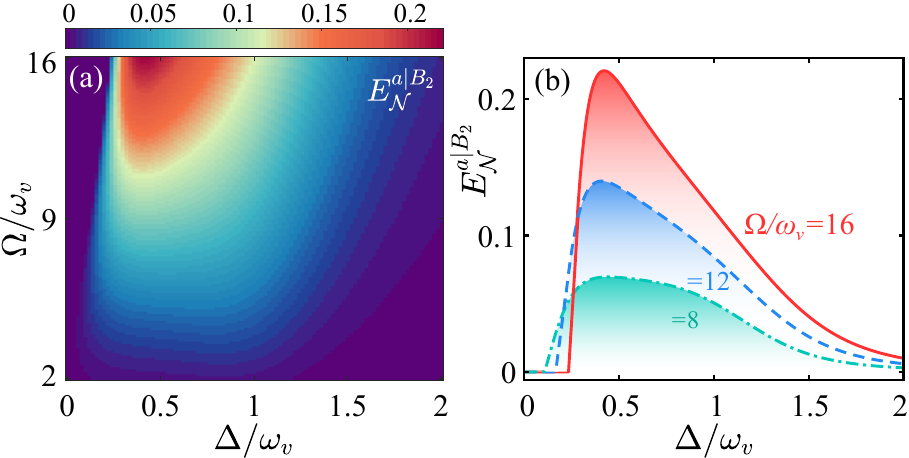}
\caption{Logarithmic negativity (a) $E_{\mathcal{N}}^{a|B_{2}}$ versus the detuning $\Delta/\omega_{v}$ and the driving amplitude $\Omega/\omega_{v}$. (b)  $E_{\mathcal{N}}^{a|B_{2}}$ versus $\Delta/\omega_{v}$ for different values of $\Omega$. Other parameters used are $\gamma_{l=1,2}/\omega_{v}=10^{-4}$, $\kappa/\omega_{v}=1/3$, $M=0$, $N=100$, and $\bar{n}=0.01$.}
\label{Fig2}
\end{figure}
%%%%%%%%%%%%%%%%%%%%%%%%%%%%%%

\emph{Generating cavity-vibration entanglement.}---To measure quantum entanglement of the system, we initially provide the definition of logarithmic negativity. For a two-mode system composed by bosonic modes $C$ and $D$, its covariance matrix $\mathcal{V}$
can be  represented as $\mathcal{V}=\{\{\mathcal{V}_{C},\mathcal{V}_{CD}\},\{\mathcal{V}_{CD}^{T}, \mathcal{V}_{D}\}\}$,
where $\mathcal{V}_{C}$, $\mathcal{V}_{CD}$, and $\mathcal{V}_{D}$ are, respectively,  the block matrices associated with the mode $C$, the correlation term, and the mode $D$.  The definition of  logarithmic negativity~\cite{Simon2000,Vidal2002,Plenio2005} is
\begin{equation}
E_{\mathcal{N}}=\max [0,-\ln(2\eta^{-})],\label{negativity}
\end{equation}
where $\eta^{-}=2^{-1/2}\{\Sigma(\mathcal{V})-[\Sigma(\mathcal{V})^{2}-4\det\mathcal{V}]^{1/2}\}^{1/2}$ is the minimum symplectic eigenvalue of the matrix $\mathcal{V}$, with $\Sigma\left(\mathcal{V}\right)=\det \mathcal{V}_{C}+\det \mathcal{V}_{D}-2\det \mathcal{V}_{CD}$. For our system, the quantum entanglement between the two modes $a$ and $B_{l}$ ($B_{1}$ and $B_{2}$) is represented by the logarithmic negativity $E_{\mathcal{N}}^{a|B_{l}}$ ($E_{\mathcal{N}}^{B_{1}|B_{2}}$).  The reduced matrix $\mathcal{V}$ of two involved modes is obtained by tracing out the uncorrelated  rows and columns in the covariance matrix $\mathbf{V}$.

Since the two types cavity-vibration entanglement ($E_{\mathcal{N}}^{a|B_{1}}$ and $E_{\mathcal{N}}^{a|B_{2}}$) exhibit similar behaviors with the change of the adjustable parameters, in Fig.~\ref{Fig2} we only show $E_{\mathcal{N}}^{a|B_{2}}$ versus the detuning $\Delta$ and the driving amplitude  $\Omega$. In the numerical simulations, we adopt parameters that are experimentally achievable: $\omega_{l}/2\pi=300$ THz, $\omega_{v}/2\pi=30$ THz, $\kappa/2\pi=10$ THz, $\gamma_{1}/2\pi=\gamma_{2}/2\pi=0.03$ THz, $g_{v}/2\pi=30$ GHz, $T\approx312$ K, $M=0$, and $N=100$. For the vibrational-mode frequency $\omega_{v}/2\pi=30$ THz,  the corresponding thermal phonon number of vibrational mode can be estimated as $\bar{n}_{l}=\bar{n}=0.01$ when $T\approx312$ K. For convenience, we take the vibrational-mode frequency $\omega_{v}$ as the scaling unit in the following numerical simulations. Figure~\ref{Fig2} shows that $E_{\mathcal{N}}^{a|B_{2}}$ reaches a peak value around the optimal detuning $\Delta/\omega_{v}\approx0.4$ when $\Omega$ takes maximum value. However, the cavity mode and the vibrational mode are uncorrelated when $\Omega$ has a small value. The reason is that the existence of  entanglement requires a large effective coupling strength $G_{2}=\sqrt{N}g_{v}\langle a\rangle_{\text{ss}}$, i.e., a strong pump field with a large amplitude $\Omega$.
Notice that we study the entanglement in the red-detuned regime, because of a large coupling strength ($|G_{2}|/\omega_{v}\approx0.31$ when $\Omega/\omega_{v}=16$ and $\Delta/\omega_{v}=0.4$) that may generate considerable entanglement~\cite{Vitali2007prl, Genes2008}. This falls into the ultrastrong coupling regime in which the counter-rotating-wave term $G_{2}(\delta a\delta B_{2}+\delta a^{\dagger}\delta B_{2}^{\dagger})$ survives. It turns out to be crucial for creating the entanglement~\cite{Huang1994}.

%%%%%%%%%%%%%%%%%%%%%%%%%%%%%
\begin{figure}[tbp]
\centering
\includegraphics[width=0.48 \textwidth]{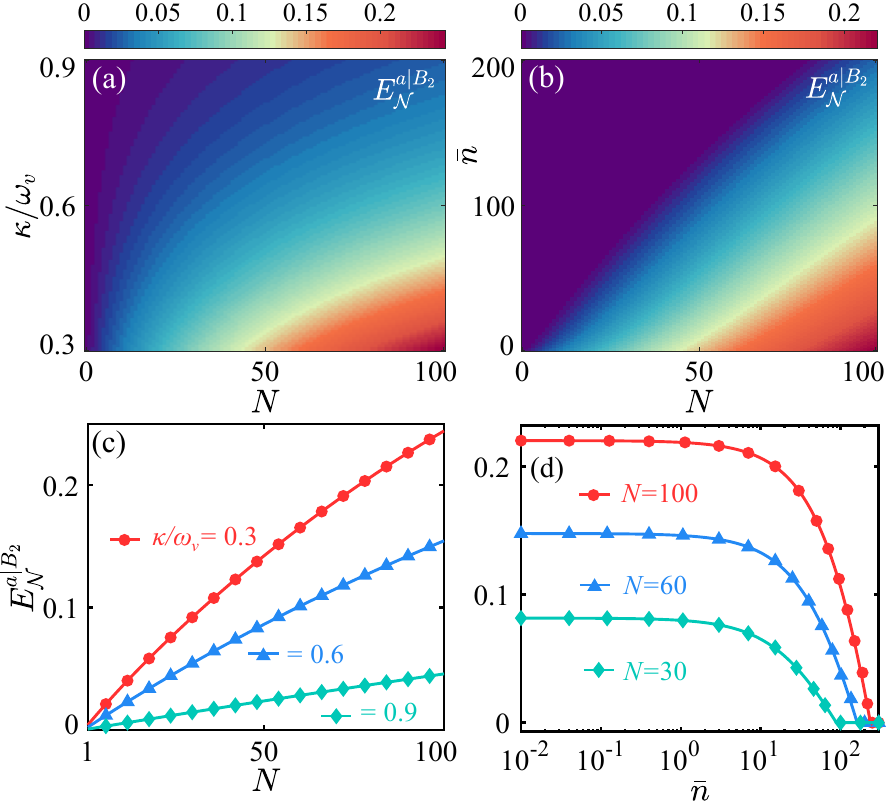}
\caption{$E_{\mathcal{N}}^{a|B_{2}}$ versus the number of molecules $N$ and the decay rate $\kappa/\omega_{v}$. (b) $E_{\mathcal{N}}^{a|B_{2}}$ versus $N$ and the thermal phonon excitation number $\bar{n}$. $E_{\mathcal{N}}^{a|B_{2}}$ versus (c) $N$  and (d) $\bar{n}$. Here $\Delta/\omega_{v}=0.4$, $\Omega/\omega_{v}=16$, and other parameters are the same as those in Fig~\ref{Fig2}.}
\label{Fig3}
\end{figure}
%%%%%%%%%%%%%%%%%%%%%%%%%%%%%%

To investigate the dependence of cavity-vibration entanglement on the collective coupling strength ($\sqrt{N}G_{v}$), in Fig.~\ref{Fig3}(a) and~\ref{Fig3}(c) we depict the $E_{\mathcal{N}}^{a|B_{2}}$ versus the number of molecules $N$ and the decay rate $\kappa$. The results show that  $E_{\mathcal{N}}^{a|B_{2}}$ increases with the number $N$,  and the maximum value of $E_{\mathcal{N}}^{a|B_{2}}$ can reach is smaller for the higher decay rates. The underlying physics for this phenomenon is that $E_{\mathcal{N}}^{a|B_{2}}$  is a monotonically increasing function of the collective coupling strength $G_{2}=\sqrt{N}G_{v}$. The increase in number $N$ can greatly enhance the $G_{2}$, thereby correspondingly enhancing the cavity-vibration entanglement. However, the decay rate $\kappa$ of cavity mode is a detrimental factor in the creation of entanglement. In Fig.~\ref{Fig3}(b), we  show the  robust quantum entanglement against the thermal noise by plotting the $E_{\mathcal{N}}^{a|B_{2}}$ as function of number $N$ and the thermal phonon excitation number $\bar{n}$. The curves in Fig.~\ref{Fig3}(d) show that $E_{\mathcal{N}}^{a|B_{2}}$ can still exist even when the thermal phonon numbers in the vibrational mode are about $\bar{n}=200$, which implies that the cavity-vibration entanglement can not only emerge in environments far above room temperature ($T\approx312$ K), but also possesses strong robustness against thermal noise.

%%%%%%%%%%%%%%%%%%%%%%%%%%%%%%
\begin{figure}[tbp]
\centering
\includegraphics[width=0.48 \textwidth]{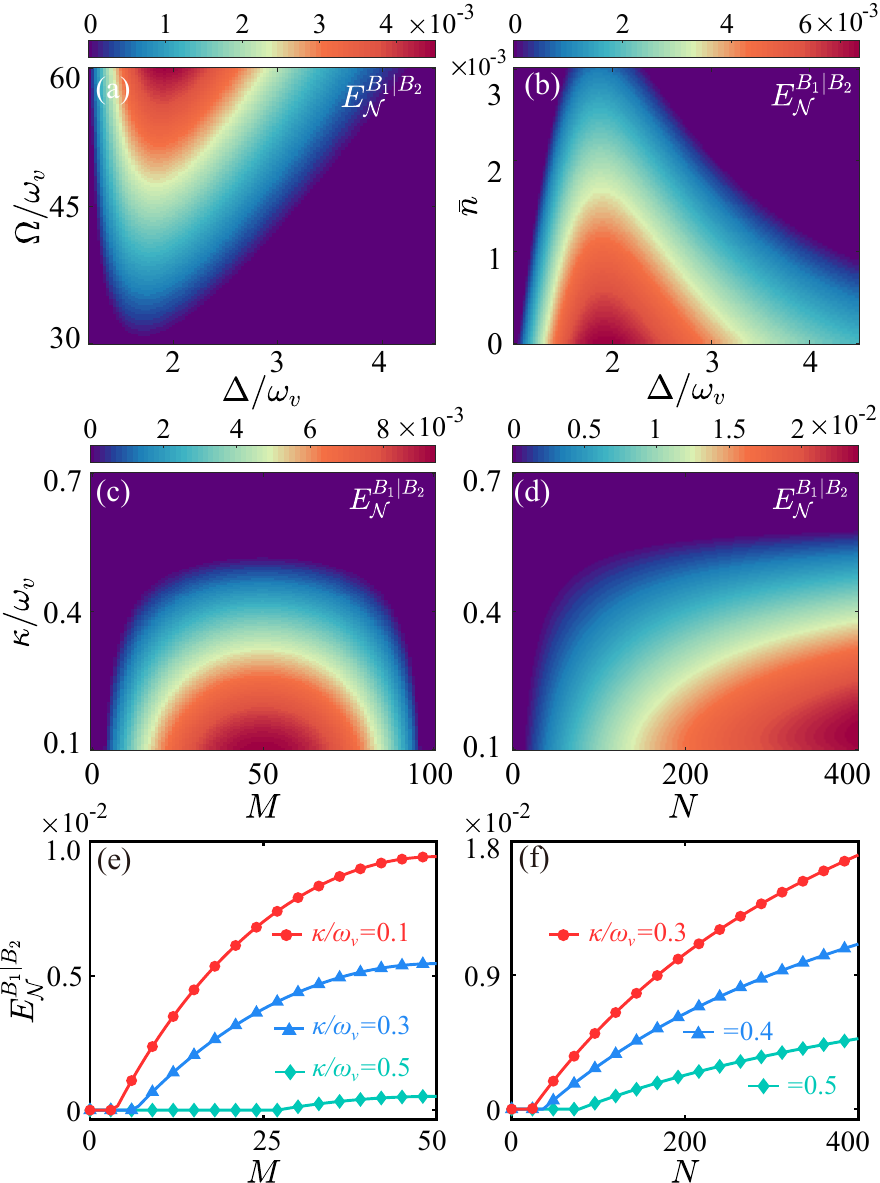}
\caption{(a) $E_{\mathcal{N}}^{B_{1}|B_{2}}$ versus the scaled detuning $\Delta/\omega_{v}$ and  $\Omega/\omega_{v}$. (b) $E_{\mathcal{N}}^{B_{1}|B_{2}}$ versus $\Delta/\omega_{v}$ and thermal phonon numbers $\bar{n}$.  (c) $E_{\mathcal{N}}^{B_{1}|B_{2}}$ versus $M$ and the scaled decay rate $\kappa/\omega_{v}$. (d) $E_{\mathcal{N}}^{B_{1}|B_{2}}$ versus the total number of molecules $N$ and $\kappa/\omega_{v}$ when $M= N/2$. $E_{\mathcal{N}}^{B_{1}|B_{2}}$ versus (e) $M$ and (f) $N$ for different values of $\kappa$. Here $\Omega/\omega_{v}=60$, $\gamma_{l}/\omega_{v}=0.3$, $\Delta/\omega_{v}=2$, $M=50$, and $\bar{n}_{1}=\bar{n}_{2}=0.001$ ($T\approx210$ K) when they are not variables, and other parameters are the same as those in Fig~\ref{Fig2}.}
\label{Fig4}
\end{figure}
%%%%%%%%%%%%%%%%%%%%%%%%%%%%%%

\emph{Generating vibration-vibration entanglement.}---Since the investigated system involves two collective vibrational modes coupled to the plasmonic cavity, a natural question emerges: can the plasmonic cavity induce quantum entanglement between the two collective vibrational modes? To clarify this issue,  it is essential to derive a reduced Hamiltonian containing only two collective vibrational modes.  In the large cavity-decay regime, by adiabatically eliminating the cavity mode~\cite{SMaterial},  the effective Hamiltonian can be expressed as
\begin{equation}
H_{\text{eff}}\approx \sum_{l=1}^{2}(\Omega _{l}-i\Gamma _{l})\delta B_{l}^{\dag}\delta B_{l}-i\mathcal{G}(\delta B_{1}^{\dag }-\delta B_{1})(\delta B_{2}^{\dag }-\delta B_{2}),
\end{equation}
where $\Omega _{l}$ and $\Gamma _{l}$ are the effective frequency and decay rate of the $l$th collective vibration mode, respectively; $\mathcal{G}\approx (i/2\omega _{v}-1/\kappa)G_{1}G_{2}$ is the effective coupling strength~\cite{SMaterial}.

In Fig.~\ref{Fig4}(a) we study the vibration-vibration entanglement by plotting $E_{\mathcal{N}}^{B_{1}|B_{2}}$ as function of $\Delta$ and $\Omega$. The results show that the optimal detuning for the peak value of $E_{\mathcal{N}}^{B_{1}|B_{2}}$ is located around $\Delta/\omega_{v}\approx2$, and a considerable driving amplitude $\Omega$ is required to generate vibration-vibration entanglement. This  phenomenon can be explained by: As the detuning $\Delta$ increases, a stronger driving amplitude is necessary to yield a larger effective coupling ($|G_{l=1,2}|/\omega_{v}\approx0.21$ when $\Omega/\omega_{v}=60$ and $\Delta/\omega_{v}=2$). We also find that the vibration-vibration entanglement is much smaller than cavity-vibration entanglement,  because there is no direct interaction between the two collective vibrational modes,  and the indirect coupling  $i\mathcal{G}$ induced by cavity mode is much smaller than $G_{1}$ and $G_{2}$. Moreover, the vibration-vibration entanglement is exceedingly susceptible to thermal noise ($E_{\mathcal{N}}^{B_{1}|B_{2}}=0$ when $\bar{n}>0.003$), as shown in Fig.~\ref{Fig4}(b).

To further investigate the dependence of vibration-vibration entanglement on the number of molecules, in Figs.~\ref{Fig4}(c) and ~\ref{Fig4}(e) we plot $E_{\mathcal{N}}^{B_{1}|B_{2}}$ versus the number $M$ (in the collective mode $B_{1}$) and decay rate $\kappa$.  The results show that $E_{\mathcal{N}}^{B_{1}|B_{2}}$ initially start at zero,  as the number $M$ increases, $E_{\mathcal{N}}^{B_{1}|B_{2}}$ correspondingly rises, reaching a peak value at the point $M=50$. Following this, $E_{\mathcal{N}}^{B_{1}|B_{2}}$ decreases back to zero in the remaining region. This phenomenon can be elucidated by the indirect coupling strength $\mathcal{G}$,  which is determined by the product of two collective optomechanical coupling  strengths ($\mathcal{G}\propto\sqrt{M}G_{v}\times \sqrt{N-M}G_{v}$). In this scenario, the induced coupling strength reaches its maximum when $M=50$, resulting in the strongest vibration-vibration entanglement. On the contrary, the entanglement becomes relatively weak and may even be completely disrupted by thermal noise when $M$ takes a larger or smaller value.

In Figs.~\ref{Fig4}(d) and ~\ref{Fig4}(f), we study the influence of total number of molecules $N$ on the vibration-vibration entanglement. Figure~\ref{Fig4}(d) shows that $E_{\mathcal{N}}^{B_{1}|B_{2}}$ grows from initial $0$ to a larger value $0.018$ with the number $N$ increases.  Because the increase of the number $N$ can significantly enhance the indirect coupling between the two collective modes, transforming them from independent to strongly entangled. Moreover, we observed from Fig~\ref{Fig4}(f) that $E_{\mathcal{N}}^{B_{1}|B_{2}}$ exhibits a approximate linear growth with respect to the total number $N$, the reason may be that the induced indirect coupling strength $\mathcal{G}$ is proportional to $N$ ($\propto NG^{2}_{v}/2$) when we set the parameter condition $M=N/2$.

Finally, the quantum entanglement can be measured by the  time-resolved detection of photons emitted from the  plasmonic cavity field. Concretely, the cavity mode quadratures can be measured by homodyne detection of the cavity output. The vibrational mode quadratures can be transferred to time-gated emitted photons, which can then be  homodyne detected through interference using an additional microwave field. The cavity homodyned signal and the molecular homodyned signal can be used to find the correlations involving field and vibration quadratures.

\emph{Conclusion.}---Our scheme of molecular optomechanics works with the red detuning. This is distinct from the blue-detuning of which the cavity magnomechanics is in favor but the molecular optomechanics may not~\cite{Li2018,Zhang2019b}. Here the blue detuning yields a narrow window for the optomechanical coupling, i.e., $G_{i}\lesssim 0.006\omega_v$ due to the stability~\cite{Vitali2007prl}. Nevertheless, the red detuning leads to $G_{i} \lesssim 0.6\omega_v$ which is much broader. Therefore the molecular optomechanical systems are in favor of the red detuning rather than the blue one, for creating strong entanglement. Our calculations reveal an entanglement $\lesssim 7\times 10^{-4}$ in the blue-detuned regime.

In summary, we developed a molecular cavity optomechanical scheme for generating the quantum entanglement in the collective vibrations of molecular ensembles. The results highlighted the role of the significant collective enhancement (i.e., $\sqrt{N}$ scaling) which yields a strong molecule-cavity entanglement at room temperature with experimentally feasible parameters. The results further revealed the quantum entanglement between the vibrational modes which shows a long-range nature and a dramatic enhancement with the number of molecules. Our scheme offers a new paradigm for the quantum interfaces between molecules and plasmonic cavities. These may form promising platforms of quantum information processing and reactivity control, where the molecular vibrations can function as the information storage medium that can be transferred to others.

\begin{acknowledgments}
We thank Deng-Gao Lai, Ya-Feng Jiao, and Jing-Xue Liu for helpful discussions. Z.Z. and J.H. gratefully acknowledge the support of the Early Career Scheme from Hong Kong Research Grants Council (No.~21302721), the National Science Foundation of China (No.~12104380), and the National Science Foundation of China/RGC Collaborative Research Scheme (No.~CRS-CUHK401/22). D. L. gratefully acknowledges the support of the National Science Foundation of China, the Excellent
Young Scientist Fund (No.~62022001). G.S.A thanks the Air Force Office of Scientific Research (Award No.~FA-9550-20-1-0366) and the Robert A. Welch Foundation (Grant No.~A-1943-20210327).
\end{acknowledgments}

%%%%%%%%%% Merge with supplemental materials %%%%%%%%%%

\onecolumngrid
\newpage
%%%%%%%%%% Prefix a "S" to all equations, figures, tables and reset the counter %%%%%%%%%%
\setcounter{equation}{0} \setcounter{figure}{0}
\setcounter{table}{0}
\setcounter{page}{1}\setcounter{secnumdepth}{3} \makeatletter
\renewcommand{\theequation}{S\arabic{equation}}
\renewcommand{\thefigure}{S\arabic{figure}}
\renewcommand{\bibnumfmt}[1]{[S#1]}
\renewcommand{\citenumfont}[1]{S#1}
\renewcommand\thesection{S\arabic{section}}
%%%%%%%%%% Prefix a "S" to all equations, figures, tables and reset the counter %%%%%%%%%%

\begin{center}
{\large \bf Supplementary Material for ``Collective Quantum Entanglement in Molecular Cavity Optomechanics"}
\end{center}

\section{Microscopic derivation of the single-photon optomechanical coupling coefficient}

The interaction Hamiltonian between the molecule including the electronic and vibrational states and the plasmonic cavity mode can be described as $(\hbar=1)$
\begin{equation}
H=\omega _{p}a^{\dagger }a+\omega _{v}b^{\dagger }b+\omega _{e}\sigma _{+}\sigma _{-}+\lambda \omega_{v}\sigma _{+}\sigma _{-}(b^{\dagger }+b) - \mu (\sigma_{+}+\sigma _{-}) E,\label{Hamilt}
\end{equation}
where $a$ $(a^{\dagger }) $ and $b$ $( b^{\dagger }) $ are, respectively, the annihilation (creation) operators of plasmonic cavity mode with resonance frequency $\omega _{p}$ and the vibrational mode with the resonance frequency $\omega _{v}$. The electronic state of the molecule is described by the raising operator $\sigma _{+}=\vert e\rangle\langle g\vert$ and lowering operator $\sigma _{-}=\vert g\rangle\langle e\vert$ with the excited state $\vert e\rangle$ and the ground state $\left\vert g\right\rangle$, and $\omega _{e}$ is the energy separation between the two states. The coupling between the electronic state and the vibrational mode is characterized by the Franck-Condon factor $\lambda$. The parameter $\mu$ is the induced Raman dipole,  which describes the interaction between the electronic state and the quantized electromagnetic field of plasmonic cavity mode $E=\sqrt{\hbar \omega _{p}/(2\varepsilon_{0}V_{p})}(a^{\dagger }+a)$, where $V_{p}$ is  the effective mode volume of plasmonic cavity and $\varepsilon _{0}$ is the vacuum permittivity.

By applying the polaron transformation  defined by the operator $U=\exp[-\lambda \sigma _{+}\sigma _{-}( b^{\dagger }-b)]$,  we can obtain
the transformed Hamiltonian
\begin{equation}
H_{1} =\omega _{p}a^{\dagger}a+\omega _{v}b^{\dagger }b+\tilde{\omega}_{e}\sigma _{+}\sigma_{-} - f \left(a^{\dagger}+a \right)\left[\sigma _{+}e^{\lambda (b^{\dagger }-b) }+\sigma _{-}e^{-\lambda ( b^{\dagger }-b)}\right],
\end{equation}
where $\tilde{\omega} _{e}= \omega _{e}-\lambda^{2}\omega _{v}$ and $f=\mu\sqrt{\hbar \omega _{p}/(2\varepsilon_{0}V_{p})}$. In the rotating frame defined by $V(t)=\exp( -iH_{0}t)$ with $H_{0} = \omega _{p}a^{\dagger}a+\omega _{v}b^{\dagger }b+ \tilde{\omega} _{e}\sigma _{+}\sigma_{-}$,  the Hamiltonian $H_{1}$ becomes
\begin{equation}
H_{2}(t) = -f \left(a^{\dagger}e^{i\omega_{p}t}+ae^{-i\omega_{p}t}\right)\left[ \sigma _{+}e^{i\tilde{\omega}_{e}t}e^{\lambda (
b^{\dagger }e^{i\omega _{v}t}-be^{-i\omega _{v}t}) }+\sigma _{-}e^{-i \tilde{\omega}_{e}t}e^{-\lambda ( b^{\dagger }e^{i\omega
_{v}t}-be^{-i\omega _{v}t}) }\right].
\label{Hamilt2t}
\end{equation}
Based on the Hamiltonian $H_{2}(t)$, we project the second-order perturbation term onto the ground state of the electronic state,
\begin{eqnarray}
&&\langle g\vert \sigma _{-}e^{-i\tilde{\omega}_{e}t}e^{-\lambda(b^{\dagger }e^{i\omega _{v}t}-b e^{-i\omega _{v}t}) }\sigma_{+}e^{i\tilde{\omega}_{e}t^{\prime }}e^{\lambda ( b^{\dagger
}e^{i\omega _{v}t^{\prime}}-be^{-i\omega _{v}t^{\prime }})
}\vert g\rangle f^{2} (a^{\dagger}e^{i\omega_{p}t}+ae^{-i\omega_{p}t})(a^{\dagger}e^{i\omega_{p}t^{\prime }}+ae^{-i\omega_{p}t^{\prime }}) \notag  \\
&=&e^{-\lambda (b^{\dagger }e^{i\omega _{v}t}-be^{-i\omega_{v}t}) }e^{\lambda ( b^{\dagger }e^{i\omega _{v}t^{\prime}}-be^{-i\omega _{v}t^{\prime }}) }e^{-i\tilde{\omega}_{e}(t-t^{\prime }) }f^{2} (a^{\dagger}e^{i\omega_{p}t}+ae^{-i\omega_{p}t})(a^{\dagger}e^{i\omega_{p}t^{\prime }}+ae^{-i\omega_{p}t^{\prime }}) \notag  \\
&\approx&[ 1+\lambda ( b^{\dagger }e^{i\omega _{v}t^{\prime}}-be^{-i\omega _{v}t^{\prime }}-b^{\dagger }e^{i\omega _{v}t}+be^{-i\omega
_{v}t})] e^{-i\tilde{\omega}_{e}( t-t^{\prime })}f^{2} (a^{\dagger}e^{i\omega_{p}t}+ae^{-i\omega_{p}t})(a^{\dagger}e^{i\omega_{p}t^{\prime }}+ae^{-i\omega_{p}t^{\prime }}),
\end{eqnarray}
then the effective Hamiltonian between the plasmonic cavity mode and the vibrational mode can be obtained as
\begin{equation}
\int_{-\infty }^{T}dt\int_{-\infty }^{t}dt^{\prime }\lambda f^{2} ( b^{\dagger }e^{i\omega _{v}t^{\prime}}-be^{-i\omega _{v}t^{\prime }}-b^{\dagger }e^{i\omega _{v}t}+be^{-i\omega
_{v}t})e^{-i\tilde{\omega}_{e}( t-t^{\prime })}(a^{\dagger}e^{i\omega_{p}t}+ae^{-i\omega_{p}t})(a^{\dagger}e^{i\omega_{p}t^{\prime }}+ae^{-i\omega_{p}t^{\prime }}). \label{effectHamilt}
\end{equation}
Referring to  Eq.~(\ref{effectHamilt}), after carrying out a series of calculations, we can derive $b^{\dagger}a^{\dagger}a$, $ba^{\dagger}a$, $b^{\dagger}aa^{\dagger}$ and $baa^{\dagger}$ terms.

On the other hand, the effective coupling  between the  cavity mode and the vibrational mode can  be rewritten as
\begin{equation}
\langle g\vert \mathcal{T}e^{-i\int_{-\infty }^{T}H_{2}(t) dt}\vert g\rangle \approx 1-i\int_{-\infty}^{T}\langle g\vert H_{2}(t) \vert g\rangle dt, \label{Veff}
\end{equation}
where $\mathcal{T}$ is the time-ordering operator. By comparing the two Eqs.~(\ref{effectHamilt}) and ~(\ref{Veff}), we can obtain the effective Hamiltonian
\begin{equation}
\langle g\vert H_{2}(t)\vert g\rangle=(g_{v}b^{\dagger }e^{i\omega _{v}t}+g_{v}^{\ast }be^{-i\omega_{v}t}) a^{\dagger }a,
\end{equation}
where we introduce the single-photon optomechanical coupling coefficient
\begin{equation}
g_{v}\approx -2\lambda f^{2} M^{-1}(v)=-2\lambda f^{2} \left[ \frac{\omega_{v}+\tilde{\omega}_{e}}{
( \omega _{v}+\tilde{\omega}_{e}) ^{2}-\omega _{p}^{2}}-\frac{\omega _{v}-\tilde{\omega}_{e}}{( \omega _{v}-\tilde{\omega}
_{e}) ^{2}-\omega _{p}^{2}}-\frac{2\tilde{\omega}_{e}}{\tilde{\omega}_{e}^{2}-\omega _{p}^{2}}\right]. \label{singlephoton}
\end{equation}
By comparing Eq.~(\ref{singlephoton}) with the expression for $g_{v}=-(\omega_{p}/\varepsilon_{0}V_{p})\sqrt{\hbar/(2m\omega _{v})}(\partial \alpha/\partial x)$ derived by phenomenological method, we can obtain the isotropic Raman tensor element
\begin{equation}
\frac{\partial \alpha}{\partial x}=\frac{2\lambda f^{2}}{M(v)}\frac{\varepsilon_{0}V_{p}}{\omega_{p}}\sqrt{\frac{2m\omega _{v}}{\hbar}}=\frac{\lambda \vert \mu \vert ^{2}}{M(v)\hbar}\sqrt{\frac{2m\omega _{v}}{\hbar}},
\end{equation}
where $\alpha$ is the  Raman polarizability and $x$ is the molecular displacement.

\section{Derivation of the effective Hamiltonian involving two collective vibrational modes}

Here we  provide a detailed derivation of the effective Hamiltonian for the two-coupled collective vibrational modes.  Specifically, in the regime of large cavity-mode decay, the three-mode molecular optomechanical system, consisting of one cavity mode and two collective vibrational modes, is reduced to a  two-coupled collective mode system by adiabatically eliminating the cavity mode. Consequently, the effective Hamiltonian for this two-mode system can be obtained. To this end,  we start from the linearized quantum Langevin equations
\begin{subequations}
\label{LangevinS}
\begin{align}
\label{LangevinSa}
\delta \dot{a}(t) =&-( i\Delta +\kappa) \delta a (t)-i\sum_{l=1}^{2}G_{l}[\delta B_{l}^{\dagger }(t)+\delta B_{l}(t)] +\sqrt{2\kappa }a_{\text{in}}(t),  \\
\label{LangevinSb}
\delta \dot{B}_{1}(t) =&-( i\omega_{1} +\gamma_{1}) \delta B_{1}(t)-iG_{1}^{\ast}\delta a(t)-iG_{1}\delta a^{\dagger}(t)+\sqrt{2\gamma}B_{1,\text{in}}(t), \\
\label{LangevinSc}
\delta \dot{B}_{2}(t)=&-(i\omega_{2} +\gamma_{2}) \delta B_{2}(t)-iG_{2}^{\ast }\delta a(t)-iG_{2}\delta a^{\dagger}(t)+\sqrt{2\gamma }B_{2,\text{in}}(t),
\end{align}
\end{subequations}
where $\omega_{l=1,2}=\omega_{v}$ is the resonance frequency for the $l$th collective vibrational mode, other parameters and variables have been defined in the main text. For convenience, below we assume that the linearized collective optomechanical coupling strengths $G_{1}$  and $G_{2}$ are real. To obtain the two-coupled vibrational mode system, we consider that the system works in the parameter regime $\omega_{l}\gg \kappa \gg \{G_{l},\gamma_{l}\}$. In this case, the solution for the fluctuation operator $\delta a(t)$ can be obtained as
\begin{eqnarray}
\delta a(t) &=&\delta a(0) e^{-( \kappa +i\Delta) t}-iG_{1}\left(\frac{1-e^{-(\kappa +i\Delta +i\omega
_{1}) t}}{\kappa +i( \Delta +\omega _{1})}\delta B_{1}^{\dag}(t) + \frac{1-e^{-( \kappa +i\Delta -i\omega
_{1}) t}}{ \kappa +i( \Delta -\omega _{1}) }\delta B_{1}( t)\right) \notag   \\
&&-iG_{2}\left( \frac{1-e^{-( \kappa+i\Delta +i\omega _{2}) t}}{ \kappa +i(\Delta +\omega
_{2}) }\delta B_{2}^{\dag }(t)+ \frac{1-e^{-\left( \kappa+i\Delta -i\omega _{2}\right) t}}{ \kappa +i( \Delta -\omega_{2}) }\delta B_{2}(t)\right) +A_{\text{in}}(t),
\end{eqnarray}
where $\delta a(0)$  is the initial value of $\delta a(t)$,  and
\begin{equation}
A_{\text{in}}(t)=\sqrt{2\kappa }e^{-(\kappa +i\Delta)t}\int_{0}^{t}a_{\text{in}}(s)e^{(\kappa +i\Delta)s}ds
\end{equation}
is the new noise operator associated with cavity mode. Here we consider the time scale $t\gg 1/\kappa$ and ignore the initial value $\delta a(0)$,  then the  solution for $\delta a(t)$ is approximated as
\begin{eqnarray}
\label{dletaat}
\delta a(t)&\approx &-\frac{iG_{1}}{\kappa +i(\Delta +\omega_{1})}\delta B_{1}^{\dagger}(t)-\frac{iG_{1}}{\kappa +i(\Delta-\omega_{1})}\delta B_{1}(t)-\frac{iG_{2}}{\kappa +i(\Delta+\omega_{2})}\delta B_{2}^{\dagger}(t)\nonumber\\
&&-\frac{iG_{2}}{\kappa +i(\Delta -\omega_{2})}\delta B_{2}(t)+A_{\text{in}}(t).
\end{eqnarray}
By substituting  Eq.~(\ref{dletaat}) into Eqs.~(\ref{LangevinSb}) and~(\ref{LangevinSc}),  the equations of motion become
\begin{subequations}
\label{fullb1b2}
\begin{align}
\label{fullb1}
\delta\dot{B}_{1}(t)=&\left(\frac{G_{1}^{2}}{\kappa-i(\Delta-\omega_{1})}-\frac{G_{1}^{2}}{\kappa +i(\Delta+\omega _{1})}\right)\delta B_{1}^{\dagger}(t)+\left(\frac{G_{1}^{2}}{\kappa-i(\Delta+\omega_{1})}-
\frac{G_{1}^{2}}{\kappa+i(\Delta-\omega_{1})}-\gamma_{1}-i\omega_{1}\right)\delta B_{1}(t)\nonumber \\
&+\left(\frac{G_{1}G_{2}}{\kappa-i(\Delta-\omega_{2})}-\frac{G_{1}G_{2}}{\kappa +i(\Delta +\omega_{2})}\right)\delta B_{2}^{\dagger }(t)+\left(\frac{G_{1}G_{2}}{\kappa-i(\Delta+\omega_{2})}-\frac{G_{1}G_{2}}{\kappa+i(\Delta-\omega_{2})}\right)\delta B_{2}(t)\nonumber \\
&-iG_{1}A_{\text{in}}(t)-iG_{1}A_{\text{in}}^{\dagger }(t)+\sqrt{2\gamma _{1}}B_{1,\text{in}}(t),\\
\label{fullb2}
\delta\dot{B}_{2}(t)=&\left(\frac{G_{1}G_{2}}{\kappa-i(\Delta-\omega_{1})}-\frac{G_{1}G_{2}}{\kappa+i(\Delta+\omega _{1})}\right)\delta B_{1}^{\dagger}(t)+\left(\frac{G_{1}G_{2}}{\kappa-i(\Delta+\omega_{1})}-\frac{G_{1}G_{2}}{\kappa+i(\Delta -\omega_{1})}\right)\delta B_{1}(t)\nonumber\\
&+\left(\frac{G_{2}^{2}}{\kappa-i(\Delta-\omega_{2})}-\frac{G_{2}^{2}}{\kappa +i(\Delta+\omega_{2})}\right)\delta B_{2}^{\dagger }(t)+\left(\frac{G_{2}^{2}}{\kappa-i(\Delta +\omega_{2})}-\frac{G_{2}^{2}}{\kappa+i(\Delta-\omega _{2})}-\gamma_{2}-i\omega_{2}\right)\delta B_{2}(t)\nonumber \\
&-iG_{2}A_{\text{in}}(t)-iG_{2}A_{\text{in}}^{\dagger}(t)+\sqrt{2\gamma_{2}}B_{2,\text{in}}(t).
\end{align}
\end{subequations}
Based on Eqs.~(\ref{fullb1b2}), we perform the rotating-wave approximation by discarding the $\delta B_{1}^{\dag }(t)$ term in Eq.~(\ref{fullb1})  and $\delta B_{2}^{\dag}(t) $ term in Eq.~(\ref{fullb2}),  then the equations become
\begin{subequations}
\label{rwab1b2}
\begin{align}
\delta\dot{B}_{1}(t)=&\left(\frac{G_{1}^{2}}{\kappa-i(\Delta+\omega_{1})}-\frac{G_{1}^{2}}{\kappa+i(\Delta-\omega_{1})}-\gamma_{1}-i\omega_{1}\right)\delta B_{1}(t)
+\left(\frac{G_{1}G_{2}}{\kappa-i(\Delta-\omega_{2})}-\frac{G_{1}G_{2}}{\kappa +i(\Delta +\omega_{2})}\right)\delta B_{2}^{\dagger }(t)\nonumber \\
&+\left(\frac{G_{1}G_{2}}{\kappa-i(\Delta+\omega_{2})}-\frac{G_{1}G_{2}}{\kappa+i(\Delta-\omega_{2})}\right)\delta B_{2}(t)-iG_{1}A_{\text{in}}(t)-iG_{1}A_{\text{in}}^{\dagger }(t)+\sqrt{2\gamma _{1}}B_{1,\text{in}}(t),\\
\delta\dot{B}_{2}(t)=&\left(\frac{G_{1}G_{2}}{\kappa-i(\Delta-\omega_{1})}-\frac{G_{1}G_{2}}{\kappa+i(\Delta+\omega _{1})}\right)\delta B_{1}^{\dagger}(t)+\left(\frac{G_{1}G_{2}}{\kappa-i(\Delta+\omega_{1})}-\frac{G_{1}G_{2}}{\kappa+i(\Delta -\omega_{1})}\right)\delta B_{1}(t)
\nonumber\\
&+\left(\frac{G_{2}^{2}}{\kappa-i(\Delta +\omega_{2})}-\frac{G_{2}^{2}}{\kappa+i(\Delta-\omega _{2})}-\gamma_{2}-i\omega_{2}\right)\delta B_{2}(t)-iG_{2}A_{\text{in}}(t)-iG_{2}A_{\text{in}}^{\dagger}(t)+\sqrt{2\gamma_{2}}B_{2,\text{in}}(t).
\end{align}
\end{subequations}
By introducing  the resonance frequency shift $\omega_{l,\text{opt}}$ and  the optical induced decay rate $\gamma_{l,\text{opt}}$ for $l$th ($l=1,2$) vibrational mode, with
\begin{subequations}
\begin{align}
\omega_{l,\text{opt}}=&\frac{G_{l}^{2}(\Delta+\omega_{l})}{\kappa^{2}+(\Delta +\omega_{l})^{2}}+\frac{G_{l}^{2}(\Delta -\omega_{l})}{\kappa^{2}+(\Delta-\omega_{l})^{2}},\\
\gamma_{l,\text{opt}}=&\frac{G_{l}^{2}\kappa}{\kappa^{2}+(\Delta-\omega_{l})^{2}}-\frac{G_{l}^{2}\kappa}{\kappa^{2}+(\Delta+\omega _{l})^{2}},
\end{align}
\end{subequations}
the  Eqs.~(\ref{rwab1b2})  can be rewritten as
\begin{subequations}
\label{Gammab1b2}
\begin{align}
\delta \dot{B}_{1}(t) = &-(\Gamma_{1}+i\Omega_{1})\delta  B_{1}(t)+\left(\frac{G_{1}G_{2}}{\kappa-i(\Delta-\omega_{2})}-\frac{G_{1}G_{2}}{\kappa +i(\Delta +\omega_{2})}\right)\delta B_{2}^{\dagger }(t) \notag \\
& +\left(\frac{G_{1}G_{2}}{\kappa -i( \Delta +\omega_{2}) }-\frac{G_{1}G_{2}}{\kappa +i(\Delta -\omega _{2}) }\right) \delta B_{2}(t)-iG_{1}A_{\text{in}}(t) -iG_{1}A_{\text{in}}^{\dag }(t) +\sqrt{2\gamma _{1}}B_{\text{1,in}}(t),\\
\delta \dot{B}_{2}(t) = &\left(\frac{G_{1}G_{2}}{\kappa-i(\Delta-\omega_{1})}-\frac{G_{1}G_{2}}{\kappa+i(\Delta+\omega _{1})}\right)\delta B_{1}^{\dagger}(t)+\left(\frac{G_{1}G_{2}}{\kappa-i(\Delta+\omega_{1})}-\frac{G_{1}G_{2}}{\kappa+i(\Delta -\omega_{1})}\right)\delta B_{1}(t) \notag \\
&-(\Gamma_{2}+i\Omega_{2}) \delta B_{2}(t) -iG_{2}A_{\text{in}}(t) -iG_{2}A_{\text{in}}^{\dag }(t) +\sqrt{2\gamma _{2}}B_{\text{2,in}}(t),
\end{align}
\end{subequations}
where $\Omega_{l}=\omega_{l}-\omega_{l,\text{opt}}$ and $\Gamma_{l}=\gamma_{l}+\gamma_{l,\text{opt}}$ are, respectively, the effective frequency and decay rate for the collective vibrational mode $B_{l}$. For simplicity, we present the effective coupling strengths between the two collective vibrational modes $B_{1}$ and $B_{2}$
\begin{subequations}
\begin{align}
\mathcal{G}_{1}=&\frac{G_{1}G_{2}[\kappa +i(\Delta -\omega _{v})]}{\kappa^{2}+(\Delta-\omega _{v})^{2}}-\frac{G_{1}G_{2}[\kappa -i(\Delta +\omega_{v})]}{\kappa ^{2}+(\Delta+\omega _{v})^{2}}, \\
\mathcal{G}_{2}=&\frac{G_{1}G_{2}[\kappa +i(\Delta +\omega _{v})]}{\kappa^{2}+(\Delta +\omega _{v})^{2}}-\frac{G_{1}G_{2}[\kappa -i(\Delta -\omega_{v})]}{\kappa ^{2}+(\Delta -\omega _{v})^{2}}.
\end{align}
\end{subequations}
Within the parameter regime $\omega_{1,2}\gg\kappa\gg G_{1,2}$ and in the resonance case $\Delta=\omega_{v}=\omega_{l=1,2}$, $\mathcal{G} _{1}$ and $\mathcal{G} _{2}$    can be approximated as
\begin{equation}
\mathcal{G} _{1}=-\mathcal{G} _{2}=\mathcal{G}\approx i\frac{G_{1}G_{2}}{2\omega_{v}}-\frac{G_{1}G_{2}}{\kappa}\label{effrctG},
\end{equation}
then the Eqs.~(\ref{Gammab1b2})  can be simplified to
\begin{subequations}
\label{simb1b2}
\begin{align}
\delta\dot{B}_{1}(t)=&-(\Gamma_{1}+i\Omega_{1})\delta B_{1}(t)+\mathcal{G}\delta B_{2}(t)-\mathcal{G}\delta B^{\dagger}_{2}(t)-iG_{1}A_{\text{in}}(t)-iG_{1}A_{\text{in}}^{\dagger}(t)+\sqrt{2\gamma_{1}}B_{1,\text{in}}(t),\\
\delta\dot{B}_{2}(t)=&-\mathcal{G}\delta B^{\dagger}_{1}(t)+\mathcal{G}\delta B_{1}(t)-(\Gamma_{2}+i\Omega_{2}) \delta B_{2}(t)-iG_{2}A_{\text{in}}(t)-iG_{2}A_{\text{in}}^{\dagger}(t)+\sqrt{2\gamma_{2}}B_{2,\text{in}}(t).
\end{align}
\end{subequations}
Based on the Eqs.~(\ref{simb1b2}), we can obtain the effective Hamiltonian as
\begin{eqnarray}
\label{Hamilteff}
H_{\text{eff}} &=&( \Omega _{1}-i\Gamma _{1}) \delta B_{1}^{\dag }\delta B_{1}+( \Omega_{2}-i\Gamma _{2}) \delta B_{2}^{\dag }\delta B_{2}-i\mathcal{G} (\delta B_{1}^{\dag}-\delta B_{1})(\delta B_{2}^{\dag }-\delta B_{2}).
\end{eqnarray}
It can be seen from  Eqs.~(\ref{effrctG}) and ~(\ref{Hamilteff}) that the coupling strength $\mathcal{G}$ between the two collective vibrational modes is determined by the product of the coupling strength between the cavity mode and each vibrational mode, i.e., $\mathcal{G}\propto G_{1}G_{2}=\sqrt{M(N-M)}G^{2}_{v}$. We also find that the coupling strength $\mathcal{G}$ is much smaller than $G_{1}$ and $G_{2}$, which means that the strength of vibration-vibration entanglement is much weaker than that of cavity-vibration entanglement.


\begin{thebibliography}{99}

%%%%molecular polariton
\bibitem{Guebrou2012}          S. Aberra Guebrou, C. Symonds, E. Homeyer, J. C. Plenet, Yu. N. Gartstein, V. M. Agranovich, and J. Bellessa, Coherent Emission from a Disordered Organic Semiconductor Induced by Strong Coupling with Surface Plasmons, Phys. Rev. Lett. \textbf{108}, 066401 (2012).

\bibitem{Spano2015}         F. C. Spano, Optical microcavities enhance the exciton coherence length and eliminate vibronic coupling in J-aggregates, J. Chem. Phys. \textbf{142}, 184707 (2015).

\bibitem{Flick2017}        J. Flick, M. Ruggenthaler, H. Appel, and A. Rubio, Atoms and molecules in cavities, from weak to strong coupling in quantum-electrodynamics (QED) chemistry, Proc. Natl. Acad. Sci. U.S.A. \textbf{114}, 3026 (2017)

\bibitem{Su2020}       R. Su, S. Ghosh, J. Wang, S. Liu, C. Diederichs, T. C. H. Liew, and Q. Xiong, Observation of exciton polariton condensation in a perovskite lattice at room temperature, Nat. Phys. \textbf{16}, 301 (2020).

  %%%%wide spectrum

\bibitem{Xiang2018}      B. Xiang, R. F. Ribeiro, A. D. Dunkelberger, J. Wang, Y. Li, B. S. Simpkins, J. C. Owrutsky, J. Yuen-Zhou, and W. Xiong, Two-dimensional infrared spectroscopy of vibrational polaritons, Proc. Natl. Acad. Sci. U.S.A. \textbf{115}, 4845 (2018).

\bibitem{Ribeiro2018}        R. F. Ribeiro, A. D. Dunkelberger, B. Xiang, W. Xiong, B. S. Simpkins, J. C. Owrutsky, and J. Yuen-Zhou, Theory for Nonlinear Spectroscopy of Vibrational Polaritons, J. Phys. Chem. Lett. \textbf{9}, 3766 (2018).

  \bibitem{Zhang2019a}       Z. D. Zhang, K. Wang, Z. Yi, M. S. Zubairy, M. O. Scully, and S. Mukamel, Polariton-assisted cooperativity of molecules in microcavities monitored by two-dimensional infrared spectroscopy, J. Phys. Chem. Lett. \textbf{10}, 4448 (2019).

  \bibitem{Zhang2023}      Z. Zhang, X. Nie, D. Lei, and S. Mukamel, Multidimensional Coherent Spectroscopy of Molecular Polaritons: Langevin Approach, Phys. Rev. Lett.  \textbf{130}, 103001 (2023).

%%%%%% exciton dynamics and reaction kinetics

\bibitem{Hutchison2012}         J. Hutchison, T. Schwartz, C. Genet, E. Devaux, and T. W. Ebbesen, Modifying chemical landscapes by coupling to vacuum fields, Angew. Chem. \textbf{124}, 1624 (2012).

 \bibitem{Coles2014}           D. M. Coles, N. Somaschi, P. Michetti, C. Clark, P. G. Lagoudakis, P. G. Savvidis, and D. G. Lidzey, Polaritonmediated energy transfer between organic dyes in a strongly coupled optical microcavity, Nat. Mater. \textbf{13}, 712 (2014).

 \bibitem{Thomas2016}       A. Thomas, J. George, A. Shalabney, M. Dryzhakov, S. J. Varma et al., Ground-state chemical reactivity under vibrational coupling to the vacuum electromagnetic field, Angew. Chem., Int. Ed. \textbf{55}, 11462 (2016).

 \bibitem{Dunkelberger2016}         A. D. Dunkelberger, B. T. Spann, K. P. Fears, B. S. Simpkins, and J. C. Owrutsky, Modified relaxation dynamics and coherent energy exchange in coupled vibration-cavity polaritons, Nat. Commun. \textbf{7}, 13504 (2016).

\bibitem{Herrera2016}       F. Herrera and F. C. Spano, Cavity-Controlled Chemistry in Molecular Ensembles, Phys. Rev. Lett. \textbf{116}, 238301 (2016).

\bibitem{Kowalewski2016}         M. Kowalewski, K. Bennett, and S. Mukamel, Cavity femtochemistry: Manipulating nonadiabatic dynamics at avoided crossings,  J. Phys. Chem. Lett. \textbf{7}, 2050 (2016).

\bibitem{Galego2016}         J. Galego, F. J. Garcia-Vidal, and J. Feist, Suppressing photochemical reactions with quantized light fields, Nat. Commun. \textbf{7}, 13841 (2016).

\bibitem{Martinez-Martinez2018}           L. Martinez-Martinez, R. Ribeiro, J. Campos-Gonzalez-Angulo, and J. Yuen-Zhou, Can ultrastrong coupling change ground-state chemical reactions? ACS Photonics \textbf{5}, 167 (2018).

\bibitem{Li2021}        X. Li, A. Mandal, and P. Huo, Cavity frequency-dependent theory for vibrational polariton chemistry, Nat. Commun. \textbf{12}, 1315 (2021).



%%%  polariton condensation,squeezed states,entanglement.

\bibitem{Aspelmeyer2014}         M. Aspelmeyer, T. J. Kippenberg, and F. Marquardt, Cavity optomechanics, Rev. Mod. Phys. \textbf{86}, 1391 (2014).

\bibitem{Metcalfe2014}       M. Metcalfe, Applications of cavity optomechanics, Appl. Phys. Rev. \textbf{1}, 031105 (2014).

%%%%%%%% a prosperous feature of several fields
%%%molecular electronics

\bibitem{Joachim2000}     C. Joachim, J. K. Gimzewski, and A. Aviram, Electronics using hybrid-molecular and mono-molecular devices,  Nature (London) \textbf{408}, 541 (2000).

\bibitem{Cuevas2010}      J. C. Cuevas and E. Scheer, \emph{Molecular Electronics: An Introduction to Theory and Experiment} (World Scientific, Singapore, 2010).

 \bibitem{Xiang2016}      D. Xiang, X. L. Wang, C. C. Jia, T. Lee, and X. F. Guo, Molecular-scale electronics: from concept to function, Chem. Rev. \textbf{116}, 4318 (2016).

%%%lasers

 \bibitem{Kena-Cohen2010}        S. Kena-Cohen and S. R. Forrest, Room-temperature polariton lasing in an organic single-crystal microcavity, Nat. Photonics \textbf{4}, 371 (2010).

%%%optomechanics

\bibitem{Shalabney2015}          A. Shalabney, J. George, J. A. Hutchison, G. Pupillo, C. Genet, and T. W. Ebbesen, Coherent coupling of molecular resonators with a microcavity mode, Nat. Commun. \textbf{6}, 5981 (2015).

\bibitem{Pino2015}       J. Pino, J. Feist, and F. J. Garcia-Vidal, Quantum theory of collective strong coupling of molecular vibrations with a microcavity mode, New J. Phys. \textbf{17}, 053040 (2015).

\bibitem{Xiang2024}         B. Xiang and W. Xiong, Molecular polaritons for chemistry, photonics and quantum Technologies, Chem. Rev. \textbf{124}, 2512 (2024)

%%%%%Molecular cavity optomechanics
\bibitem{Roelli2016}           P. Roelli, C. Galland, N. Piro, and T. J. Kippenberg, Molecular cavity optomechanics as a theory of plasmon-enhanced Raman scattering, Nat. Nanotechnol. \textbf{11}, 164 (2016).

\bibitem{Schmidt2016}        M. K. Schmidt, R. Esteban, A. Gonzalez-Tudela, G. Giedke, and J. Aizpurua, Quantum mechanical description of Raman scattering from molecules in plasmonic cavities, ACS Nano. \textbf{10}, 6291 (2016).

\bibitem{Benz2016}      F. Benz \emph{et al.}, Single-molecule optomechanics in ``picocavities'', Science \textbf{354}, 726 (2016).

\bibitem{Dezfouli2017}     M. K. Dezfouli and S. Hughes, Quantum optics model of surface-enhanced Raman spectroscopy for arbitrarily shaped plasmonic resonators, ACS Photonics \textbf{4}, 1245 (2017).

\bibitem{Lombardi2018}     A. Lombardi, M. K. Schmidt, L. Weller, W. M. Deacon, F. Benz, B. de Nijs, J. Aizpurua, and J. J. Baumberg, Pulsed Molecular Optomechanics in Plasmonic Nanocavities: From Nonlinear Vibrational Instabilities to Bond-Breaking, Phys. Rev. X \textbf{8}, 011016 (2018).

\bibitem{Neuman2019}      T. Neuman, R. Esteban, G. Giedke, M. K. Schmidt, and J. Aizpurua, Quantum description of surface-enhanced resonant Raman scattering within a hybrid-optomechanical model, Phys. Rev. A \textbf{100}, 043422 (2019).

\bibitem{Zhang2020}           Y. Zhang, J. Aizpurua, and R. Esteban, Optomechanical collective effects in surface-enhanced Raman scattering from
many molecules, ACS Photonics \textbf{7}, 1676 (2020)

\bibitem{Esteban2022}       R. Esteban, J. J. Baumberg, and J. Aizpurua, Molecular optomechanics approach to surface-enhanced Raman scattering, Acc. Chem. Res. \textbf{55}, 1889 (2022).

\bibitem{Jakob2023}      L. A. Jakob, \emph{et al.}, Giant optomechanical spring effect in plasmonic nano- and picocavities probed by surface-enhanced Raman scattering, Nat. Commun. \textbf{14}, 3291 (2023).

\bibitem{Koner2023}        A. Koner, M. Du, S. Pannir-Sivajothi, R. H. Goldsmith, and J. Yuen-Zhou, A path towards single molecule vibrational strong
coupling in a Fabry-Perot microcavity, Chem. Sci.  \textbf{14}, 7753 (2023)

%%%%%%%numerous physical systems
%%%quantum liquid
\bibitem{Zhou2017}      Y. Zhou, K. Kanoda, and T.-K. Ng, Quantum spin liquid states, Rev. Mod. Phys. \textbf{89}, 025003 (2017).

\bibitem{Savary2017}       L. Savary and L. Balents, Quantum spin liquids: a review, Rep. Prog. Phys. \textbf{80}, 016502 (2017).

%%%ferromagnetic materials
 \bibitem{Ghosh2003}      S. Ghosh, T. F. Rosenbaum, G. Aeppli, and S. N. Coppersmith, Entangled quantum state of magnetic dipoles, Nature (London) \textbf{425}, 48 (2003).

 \bibitem{Li2018}       J. Li, S.-Y. Zhu, and G. S. Agarwal, Magnon-Photon-Phonon Entanglement in Cavity Magnomechanics, Phys. Rev. Lett. \textbf{121},
203601 (2018).

 \bibitem{Zhang2019b}       Z. Zhang, M. O. Scully, and G. S. Agarwal, Quantum entanglement between two magnon modes via Kerr nonlinearity driven far from equilibrium, Phys. Rev. Research \textbf{1}, 023021 (2019).

%%%trapped ions
 \bibitem{Leibfried2003}        D. Leibfried, R. Blatt, C. Monroe, and D. Wineland, Quantum dynamics of single trapped ions, Rev. Mod. Phys. \textbf{75}, 281 (2003).



%%%cooling

\bibitem{Wilson-Rae2007}      I. Wilson-Rae, N. Nooshi, W. Zwerger, and T. J. Kippenberg, Theory of Ground State Cooling of a Mechanical Oscillator Using Dynamical Backaction, Phys. Rev. Lett. \textbf{99}, 093901 (2007).

\bibitem{Marquardt2007}      F. Marquardt, J. P. Chen, A. A. Clerk, and S. M. Girvin, Quantum Theory of Cavity-Assisted Sideband Cooling of Mechanical Motion, Phys. Rev. Lett. \textbf{99}, 093902 (2007).

\bibitem{Chan2011}    J. Chan, T. P. Alegre, A. H. Safavi-Naeini, J. T. Hill, A. Krause, S. Groeblacher, M. Aspelmeyer, and O. Painter, Laser cooling of a nanomechanical oscillator into its quantum ground state, Nature (London) \textbf{478}, 89 (2011).

\bibitem{Teufel2011a}    J. D. Teufel, T. Donner, D. Li, J. W. Harlow, M. S. Allman, K. Cicak, A. J. Sirois, J. D. Whittaker, K. W. Lehnert, and R. W. Simmonds, Sideband cooling of micromechanical motion to the quantum ground state, Nature (London) \textbf{475}, 359 (2011).

\bibitem{Barzanjeh2022}      S. Barzanjeh, A. Xuereb, S. Gr\"{o}blacher, M. Paternostro, C. A. Regal, and E. M. Weig, Optomechanics for quantum technologies, Nat. Phys. \textbf{18}, 15 (2022).


%%%quantum entanglement in optomechanical system

\bibitem{Mancini2002}      S. Mancini, V. Giovannetti, D. Vitali, and P. Tombesi, Entangling Macroscopic Oscillators Exploiting Radiation Pressure, Phys. Rev. Lett. \textbf{88}, 120401 (2002).

\bibitem{Pirandola2006} S. Pirandola, D. Vitali, P. Tombesi, and S. Lloyd, Macroscopic Entanglement by Entanglement Swapping, Phys. Rev. Lett. \textbf{97}, 150403 (2006).

\bibitem{Vitali2007prl}       D. Vitali, S. Gigan, A. Ferreira, H. R. B\"{o}hm, P. Tombesi, A. Guerreiro, V. Vedral, A. Zeilinger, and M. Aspelmeyer, Optomechanical Entanglement between a Movable Mirror and a Cavity Field, Phys. Rev. Lett. \textbf{98}, 030405 (2007).

\bibitem{Vitali2007jpa}      D. Vitali, S. Mancini, and P. Tombesi, Stationary entanglement between two movable mirrors in a classically driven Fabry-Perot cavity, J. Phys. A: Math. Theor. \textbf{40}, 8055 (2007).

\bibitem{Paternostro2007}    M. Paternostro, D. Vitali, S. Gigan, M. S. Kim, C. Brukner, J. Eisert, and M. Aspelmeyer, Creating and Probing Multipartite Macroscopic Entanglement with Light, Phys. Rev. Lett. \textbf{99}, 250401 (2007).

\bibitem{Genes2008}     C. Genes, A. Mari, P. Tombesi, and D. Vitali, Robust entanglement of a micromechanical resonator with output optical
fields, Phys. Rev. A \textbf{78}, 032316 (2008).

\bibitem{Hartmann2008}      M. J. Hartmann and M. B. Plenio, Steady State Entanglement in the Mechanical Vibrations of Two Dielectric Membranes, Phys. Rev. Lett. \textbf{101}, 200503 (2008).

\bibitem{Borkje2011} K. B{\o}rkje, A. Nunnenkamp, and S. M. Girvin, Proposal for Entangling Remote Micromechanical Oscillators via Optical Measurements, Phys. Rev. Lett. \textbf{107}, 123601 (2011).

\bibitem{Abdi2012}  M. Abdi, S. Pirandola, P. Tombesi, and D. Vitali, Entanglement Swapping with Local Certification: Application to Remote Micromechanical Resonators, Phys. Rev. Lett. \textbf{109}, 143601 (2012).

\bibitem{Tian2013}     L. Tian, Robust Photon Entanglement via Quantum Interference in Optomechanical Interfaces, Phys. Rev. Lett. \textbf{110}, 233602 (2013).

\bibitem{Wang2013}    Y.-D. Wang and A. A. Clerk, Reservoir-Engineered Entanglement in Optomechanical Systems, Phys. Rev. Lett. \textbf{110}, 253601 (2013).

\bibitem{Palomaki2013}      T. A. Palomaki, J. D. Teufel, R. W. Simmonds, and K. W. Lehnert, Entangling mechanical motion with microwave fields, Science \textbf{342}, 710 (2013).

\bibitem{Flayac2014}     H. Flayac and V. Savona, Heralded Preparation and Readout of Entangled Phonons in a Photonic Crystal Cavity, Phys. Rev. Lett. \textbf{113}, 143603 (2014).

\bibitem{Liao2014}   J.-Q. Liao, Q.-Q. Wu, and F. Nori, Entangling two macroscopic mechanical mirrors in a two-cavity optomechanical system, Phys. Rev. A \textbf{89}, 014302 (2014).

\bibitem{Ho2018}       M. Ho, E. Oudot, J.-D. Bancal, and N. Sangouard, Witnessing Optomechanical Entanglement with Photon Counting, Phys. Rev. Lett. \textbf{121}, 023602 (2018).

\bibitem{Barzanjeh2019}        S. Barzanjeh, E. S. Redchenko, M. Peruzzo, M. Wulf, D. P. Lewis, G. Arnold, and J. M. Fink, Stationary entangled radiation from micromechanical motion, Nature (London) \textbf{570}, 480 (2019).

\bibitem{Jiao2020}     Y.-F. Jiao, S.-D. Zhang, Y.-L. Zhang, A. Miranowicz, L.-M. Kuang, and H. Jing, Nonreciprocal Optomechanical Entanglement against Backscattering Losses, Phys. Rev. Lett. \textbf{125}, 143605 (2020).

\bibitem{Lai2022}      D.-G. Lai, J.-Q. Liao, A. Miranowicz, and F. Nori, NoiseTolerant Optomechanical Entanglement via Synthetic Magnetism, Phys. Rev. Lett. \textbf{129}, 063602 (2022).

\bibitem{Huang2022}      J. Huang, D.-G. Lai, and J.-Q. Liao, Thermal-noise-resistant optomechanical entanglement via general dark-mode control,   Phys. Rev. A \textbf{106}, 063506 (2022).

%%%macroscopic mechanical entanglement

\bibitem{Riedinger2018}     R. Riedinger, A. Wallucks, I. Marinkovi\'{c}, C. L\"{o}schnauer, M. Aspelmeyer, S. Hong, and S. Gr\"{o}blacher, Remote quantum entanglement between two micromechanical oscillators, Nature (London) \textbf{556}, 473 (2018).

\bibitem{Ockeloen-Korppi2018}       C. F. Ockeloen-Korppi, E. Damsk\"{a}gg, J.-M. Pirkkalainen,  M. Asjad, A. A. Clerk, F. Massel, M. J. Woolley, and M. A. Sillanp\"{a}\"{a}, Stabilized entanglement of massive mechanical oscillators, Nature (London) \textbf{556}, 478 (2018).

\bibitem{Kotler2021}     S. Kotler, G. A. Peterson, E. Shojaee, F. Lecocq, K. Cicak, A. Kwiatkowski, S. Geller, S. Glancy, E. Knill, R. W. Simmonds, J. Aumentado, and J. D. Teufel, Direct observation of deterministic macroscopic entanglement, Science \textbf{372}, 622 (2021).

\bibitem{Lepinay2021}     L. M. de L\'{e}pinay, C. F. Ockeloen-Korppi, M. J. Woolley, and M. A. Sillanp\"{a}\"{a}, Quantum mechanics-free subsystem with mechanical oscillators, Science \textbf{372}, 625 (2021).

%%%%%Frequency Upconversion

\bibitem{Chen2021}       W. Chen, P. Roelli, H. Hu, S. Verlekar, S. P. Amirtharaj, A. I. Barreda, T. J. Kippenberg, M. Kovylina, E. Verhagen, A. Martínez, and C. Galland, Continuous-wave frequency upconversion with a molecular optomechanical nanocavity, Science \textbf{374}, 1264 (2021).

\bibitem{Xomalis2021}        A. Xomalis, X. Zheng, R. Chikkaraddy, Z. Koczor-Benda, E. Miele, E. Rosta, G. A. E. Vandenbosch, A. Martínez, and J. J. Baumberg, Detecting mid-infrared light by molecular frequency upconversion in dual-wavelength nanoantennas, Science \textbf{374}, 1268 (2021).

%%%%NNNN
\bibitem{Agarwal1984}         G. S. Agarwal, Vacuum-Field Rabi Splittings in Microwave Absorption by Rydberg Atoms in a Cavity, Phys. Rev. Lett. \textbf{53}, 1732 (1984).

\bibitem{Zou2024}    F. Zou, L. Du, Y. Li, and H. Dong, Amplifying Frequency Up-Converted Infrared Signals with a Molecular Optomechanical Cavity, Phys. Rev. Lett. \textbf{132}, 153602 (2024).

%%%%realized in experiment

\bibitem{Mubeen2012}     S. Mubeen, S. Zhang, N. Kim, S. Lee, S. Kramer, H. Xu, and M. Moskovits, Plasmonic properties of gold nanoparticles separated from a gold mirror by an ultrathin oxide, Nano Lett. \textbf{12}, 2088 (2012).

\bibitem{Zhang2013}        R. Zhang, Y. Zhang, Z. Dong, S. Jiang, C. Zhang, L. Chen, L. Zhang, Y. Liao, J. Aizpurua, Y. Luo, J. Yang, and J. Hou, Chemical mapping of a single molecule by plasmon-enhanced Raman scattering, Nature (London) \textbf{498}, 82 (2013).


\bibitem{SMaterial}       See Supplemental Material for detailed derivation of the single-photon optomechanical coupling coefficient and the effective Hamiltonian  involving two collective vibrational modes.


%%%collective operators


\bibitem{Sun2003}      C. P. Sun, Y. Li, and X. F. Liu, Quasi-Spin-Wave Quantum Memories with a Dynamical Symmetry, Phys. Rev. Lett. \textbf{91}, 147903 (2003).

\bibitem{Emary2003}       C. Emary and T. Brandes, Quantum Chaos Triggered by Precursors of a Quantum Phase Transition: The Dicke Model, Phys. Rev. Lett. \textbf{90}, 044101 (2003).


%%%%Quantum Optics

\bibitem{Gardiner2013}      C. W. Gardiner and P. Zoller, \emph{Quantum Noise} (Springer, Berlin, 2000).


%%%Routh-Hurwitz criterion

\bibitem{Gradstein2014}     I. S. Gradshteyn and I. M. Ryzhik, \emph{Table of Integrals, Series, and Products} (Academic, New York, 2014).

%%%%logarithmic negativity


\bibitem{Vidal2002}     G. Vidal and R. F. Werner, Computable measure of entanglement, Phys. Rev. A \textbf{65}, 032314 (2002).


\bibitem{Plenio2005}   M. B. Plenio, Logarithmic Negativity: A Full Entanglement Monotone That is not Convex, Phys. Rev. Lett. \textbf{95}, 090503 (2005).

\bibitem{Simon2000}     R. Simon, Peres-Horodecki Separability Criterion for Continuous Variable Systems, Phys. Rev. Lett. \textbf{84}, 2726 (2000).

\bibitem{Huang1994}           H. Huang and G. S. Agarwal, General linear transformations and entangled states, Phys. Rev. A \textbf{49}, 52 (1994).


\end{thebibliography}
\end{document}